\documentclass[12pt]{article}
\usepackage{amssymb,amsmath,epsfig}

\begin{document}

\title{\bf Kaluza-Klein Cosmology With Modified Holographic Dark Energy}
\author{M. Sharif \thanks {msharif@math.pu.edu.pk} and Farida
Khanum\thanks {faridakhanum.math@yahoo.com}\\
Department of Mathematics, University of the Punjab,\\
Quaid-e-Azam Campus, Lahore-54590, Pakistan.}

\date{}

\maketitle
\begin{abstract}
We investigate the compact Kaluza-Klein cosmology in which modified
holographic dark energy is interacting with dark matter. Using this
scenario, we evaluate equation of state parameter as well as
equation of evolution of the modified holographic dark energy.
Further, it is shown that the generalized second law of
thermodynamics holds without any constraint.
\end{abstract}

\section{Introduction}

Recent cosmological observations \cite{2}-\cite{4} indicate that the
universe is spatially flat and has an accelerated expansion. These
observations lead to a matter called dark energy (DE) which has
large negative pressure. The DE can be explained in terms of
cosmological constant, acts like a perfect fluid with an equation of
state, satisfying the observational data so far. However, this
involves the problems of fine tuning and cosmic coincidence. Many
dynamical models like phantom \cite{5}, quintessence \cite{6},
quintom \cite{7}, tachyon \cite{8}, generalized Chaplying gas
\cite{9} etc. have been proposed to alleviate these problems. The
nature of DE is still unknown in spite of many efforts for its
investigation.

An alternative way to understand the early time inflation and late
time acceleration of the universe is to modify gravity theories such
as $f(R)$ theory, $f(T)$ theory, $f(G)$ theory and extra dimensional
theories. Recently, the theory of extra dimensions has attracted
many people. The world may have five dimensions is due to the idea
of Kaluza \cite{13} and Klein \cite{14}. They used one extra
dimension to unify gravity and electromagnetism and obtained a $5D$
general relativity. This idea has been used by many people for
studying the models of cosmology and particle physics
\cite{16}-\cite{18} (see review of the KK and higher dimensional
unified theories \cite{16a}).

In cosmology of $5D$ with pure geometry in non-compact KK theory,
one does not need to insert matter by hand because the matter is
induced in $4D$ by $5D$ vacuum theory \cite{29}. Actually, the
curvature of $5D$ spacetime induces effective properties of matter
in $4D$. This is a consequence of the Campbells theorem
\cite{29a}-\cite{29b} which states that any analytic $N$-dimensional
Riemannian manifold can be locally embedded in $N+1$-dimensional
Ricci flat Riemannian manifold. In this theory, the energy density
of scalar field contributes to define the early inflation and late
time acceleration which provides extra motivation in solving these
problems \cite{29c}. People \cite{29d}-\cite{29f} explored this $5D$
theory by inserting matter instead of pure $5D$ geometry.

The nature of DE can also be studied according to some basic quantum
gravitational principles, for example, holographic DE principle.
According to this principle \cite{32}, the degrees of freedom in a
bounded system should be finite and do not scale by its volume but
with its boundary area. Cohen et al. \cite{33} found that for a
system with infrared (long distance) cutoff scale $L$ and
ultraviolet (short distance) cutoff scale $\Lambda$ without decaying
into a black hole, the quantum vacuum energy should be less than or
equal to the mass of a black hole, i.e., $L^{3}\rho_{\Lambda}\leq
LM^{2}_{p}$. Here $\rho_{\Lambda}$ is the vacuum energy density and
$M_{p}=(8\pi G)^{-\frac{1}{2}}$ is the reduced Plank mass. Using
this idea in cosmology, one can take $L$ which satisfies this
inequality with $\rho_{\Lambda}$ as DE density.

There exist many cosmological versions of holographic principle in
literature \cite{34}-\cite{36}. It is found \cite{36} that this
principle can be replaced by the generalized second law of
thermodynamics (GSLT) for time dependent backgrounds. This is
similar to the cosmological holographic principle given by Fischler
and Susskind \cite{34} for an isotropic open and flat universe with
fixed equation of state. It is found that cold dark matter (CDM) is
decaying into DE \cite{36a,36b} which favors the interaction between
these two components. Many models with interacting DE have been
investigated \cite{42}-\cite{43}. In a recent paper \cite{36c}, we
have investigated the validity of GSLT for Bianchi type $I$ model in
which anisotropic dark energy is interacting with dark matter and
anisotropic radiation. The holographic DE in extra dimension can be
studied with the help of the mass of black hole in $N+1$ dimensional
spacetime \cite{37a} and modified holographic dark energy (MHDE)
\cite{37b}. Liu et al. \cite{38} has obtained some interesting
results with MHDE in Dvali-Gabadaze Porrati brane world.

In this paper, we study equations of state parameter as well as
evolution of MHDE in compact KK theory such that MHDE is interacting
with DM to explore the dynamics of vacuum energy. Also, we
investigate the GSLT in this scenario. The format of the the paper
is as follows: In next section, the equation of state parameter and
equation of evolution are formulated. Section \textbf{3} discusses
the validity of GSLT. Finally, we give some concluding remarks in
the last section.

\section{Equations of State Parameter and Evolution}

Here we take the modified holographic dark energy interacting with
dark matter in compact Kaluza-Klein cosmology. We explore equation
of state parameter as well as equation of evolution in this
scenario. The metric representation of the KK universe \cite{44} is
given by
\begin{equation}\label{1}
ds^{2}=dt^{2}-a^{2}(t)[\frac{dr^{2}}{1-kr^{2}}
+r^{2}(d\theta^{2}+\sin\theta^{2}d\phi^{2})+(1-kr^{2})d\psi^{2}],
\end{equation}
where a(t) is the scale factor, $k=-1,0,1$ is the curvature
parameter for the closed, flat and open universe respectively.
Suppose that the KK universe is filled with perfect fluid defined by
the following energy-momentum tensor
\begin{equation}\label{2}
T_{\mu\nu}=(P+\rho)U_{\mu}U_{\nu}-g_{\mu\nu}P,\quad(\mu,~\nu=0,1,2,3,4),
\end{equation}
where $P=P_{\Lambda}+P_{m},~\rho=\rho_{\Lambda}+\rho_m$ are the
pressure and density respectively. The subscripts $\Lambda$ and $m$
denote MHDE and DM respectively. Here $U_{\mu}$ is the five velocity
such that $U^{\mu}U_{\mu}=1$. The Einstein field equations are
\begin{equation}\label{3}
R_{\mu\nu}-\frac{1}{2}g_{\mu\nu}R=\kappa T_{\mu\nu},
\end{equation}
where $R_{\mu\nu},~g_{\mu\nu},~R,~T_{\mu\nu}$ and $\kappa$ are the
Ricci tensor, the metric tensor, the Ricci scalar, the
energy-momentum tensor and the coupling constant respectively. For
the sake of simplicity, we take $\kappa=1$. Using Eqs.(\ref{1}) and
(\ref{2}) in Eq.(\ref{3}), it follows that
\begin{eqnarray}\label{4}
\rho&=& 6\frac{\dot{a}^{2}}{a^{2}}+6\frac{k}{a^{2}},\\\label{5}
P&=&-3\frac{\ddot{a}}{a}-3\frac{\dot{a}^{2}}{a^{2}}-3\frac{k}{a^{2}}.
\end{eqnarray}
For the flat universe $k=0$, we have
\begin{eqnarray}\label{4a}
\rho= 6\frac{\dot{a}^{2}}{a^{2}}=6H^{2},\\\label{5a}
P=-3\frac{\ddot{a}}{a}-3\frac{\dot{a}^{2}}{a^{2}},
\end{eqnarray}
where $H=\frac{\dot{a}}{a}$ is the Hubble parameter. The continuity
equation, $T^{\mu\nu};_{\nu}=0$, gives
\begin{equation}\nonumber
\dot{\rho}+4H(\rho+P)=0.
\end{equation}

Now we assume that MHDE is interacting with the DM so that the
continuity equations for DM and MHDE, respectively, take the form
\begin{eqnarray}\label{6}
\dot{\rho}_{m}+4H(\rho_{m}+P_{m})=Q_{4},\quad
\dot{\rho}_{\Lambda}+4H(\rho_{\Lambda}+P_{\Lambda})=-Q_{4}.
\end{eqnarray}
Here $Q_4$ is a new form of interacting term defined as \cite{45}
\begin{equation}\label{0}
Q_{4}=3b(\rho_{\Lambda}-\rho_{m}),
\end{equation}
where $b$ is a coupling constant. In a recent paper \cite{46}, Cai
and Su found that $Q$ (an interacting term) may cross the
non-interacting line ($Q=0$), i.e., the sign of $Q$ changes around
the red-shift variable $z=0.5$. This leads to a big challenge for
the interacting models as the general interaction terms cannot
change the signs. However, this interaction term not only solves
this problem but also consistent with GSLT using the arguments
\cite{45} both for the early time when $T_{m}>T_{d}$ and for the
late time when $T_{d}>T_{m}$ ($T_m$ and $T_d$ indicate temperature
or dark matter and dark energy respectively).

In $N+1$ dimensional spacetime, the mass of the Schwarzschild black
hole is given by \cite{37a}
\begin{equation*}\nonumber
M=\frac{(N-1)A_{N-1}r^{N-2}_{H}}{16\pi G},
\end{equation*}
where $A_{N-1}$ denotes the area of unit $N$-sphere, $r_{H}$ is the
horizon scale of black hole, $G$ is $N+1$ dimensional gravitational
constant which is related by $N+1$ dimensional Plank mass $M_{N+1}$
and the usual Plank mass in $4$-dimensional spacetime. Also, $8\pi
G=M^{-(N-1)}_{N+1}=\frac{V_{N-3}}{M^{2}_{p}},~V_{N-3}$ is the volume
of extra dimensional space. Thus
\begin{equation*}\nonumber
M=\frac{(N-1)A_{N-1}r^{N-2}_{H}M^{2}_{p}}{2V_{N-3}}.
\end{equation*}
It is given that \cite{37b}
\begin{equation*}
L^{3}\rho_{\Lambda}\sim
\frac{(N-1)A_{N-1}L^{N-2}M^{2}_{p}}{2V_{N-3}}
\end{equation*}
implying that
\begin{equation*}
\rho_{\Lambda}=\frac{c^{2}(N-1)A_{N-1}L^{N-5}M^{2}_{p}}{2V_{N-3}},
\end{equation*}
where $c$ is a constant parameter. For KK cosmology, i.e., for
$N=4$, it follows that
\begin{equation}\label{8}
\rho_{\Lambda}=\frac{3c^{2}A_{3}L^{-1}}{2}.
\end{equation}
Inserting the value of area of $4$-sphere, we have
\begin{equation}\label{8a}
\rho_{\Lambda}=3c^{2}\pi^{2}L^{2}.
\end{equation}

The dynamical apparent horizon (a marginally trapped surface with
zero expansion) is defined as \cite{37c}
\begin{equation}\label{8b}
h^{ab}\partial_{a}\widetilde{r}\partial_{b}\widetilde{r}=0,\quad(a,b=0,1).
\end{equation}
Using the FRW model, it has been proved that the radius of apparent
horizon is given as
\begin{equation}\label{8c}
r_{a}=\frac{1}{\sqrt{H^{2}+\frac{k}{a^{2}}}},
\end{equation}
where $\widetilde{r}=a(t)r,~x^{0}=t,~x^{1}=r$ and $h_{ab}=diag
(-1,\frac{a^{2}}{\sqrt{1-kr^{2}}})$. It has also been proved that
the apparent horizon coincides with the Hubble horizon
$r_{H}=\frac{1}{H}$ for the flat FRW universe. In this paper, we
take KK cosmology containing FRW as subspace with compact fifth
dimension. This model has the same treatment of apparent horizon as
in the case of flat FRW model \cite{37c}. Thus we can write
\begin{equation}\label{8d}
r_{a}=\frac{1}{H}=r_{H}=L.
\end{equation}
Therefore, the infrared cutoff of the universe $L$ in this KK flat
universe is equal to the apparent horizon which coincides with
Hubble horizon.

The equations of state are
\begin{equation}\label{10}
P_{\Lambda}=\omega_{\Lambda}\rho_{\Lambda},\quad
P_{m}=\omega_{m}\rho_{m}.
\end{equation}
The continuity equations (\ref{6}) in effective theory are
\begin{eqnarray}\label{6a}
\dot{\rho}_{m}+4H(1+\omega^{eff}_{m})\rho_{m}&=&0,\\\label{7a}
\dot{\rho}_{\Lambda}+4H(1+\omega^{eff}_{\Lambda})\rho_{\Lambda}&=&0,
\end{eqnarray}
where
\begin{equation}\label{11a}
\omega^{eff}_{m}=\omega_{m}-\frac{Q_{4}}{4H\rho_{m}},\quad
\omega^{eff}_{\Lambda}=\omega_{\Lambda}+\frac{Q_{4}}{4H\rho_{\Lambda}}.
\end{equation}
Using Eqs.(\ref{8a}), (\ref{8d}) and (\ref{7a}), we have
\begin{equation}\label{11b}
\omega^{eff}_{\Lambda}=-1-\frac{\dot{\rho_\Lambda}}{4\rho_{\Lambda}}
=-1-\frac{\dot{H}}{2H^{2}}.
\end{equation}
It follows from Eq.(\ref{11a}) that
\begin{equation*}
\omega_{\Lambda}=\omega^{eff}_{\Lambda}-\frac{Q_{4}}{4H\rho_{\Lambda}}
\end{equation*}

Inserting the values of $Q_4$ and $\omega^{eff}_{\Lambda}$ from
Eqs.(\ref{0}) and (\ref{11b}) respectively, we obtain
\begin{equation}\label{13}
\omega_{\Lambda}=-1-\frac{\dot{H}}{2H^{2}}-\frac{3b(\rho_{\Lambda}-\rho_{m})}{4\rho_{\Lambda}}.
\end{equation}
Equations (\ref{4a}) and (\ref{5a}) yield
\begin{equation}\nonumber
\frac{\dot{H}}{2H^{2}}=-\frac{1}{2}(1-\frac{\ddot{a}a}{\dot{a}^{2}}).
\end{equation}
Substituting this value of $\frac{\dot{H}}{2H^{2}}$ in
Eq.(\ref{13}), we have
\begin{equation}\label{14}
\omega_{\Lambda}=-\frac{1}{2}+\frac{\ddot{a}a}{2\dot{a}^{2}}
-\frac{3b(\Omega_{\Lambda}-\Omega_{m})}{4\Omega_\Lambda},
\end{equation}
where
$\Omega_{\Lambda}=\frac{\rho_{\Lambda}}{\rho_{c}},\quad\rho_{c}=6H^{2}$.
This is the equation of state parameter for the modified holographic
dark energy.

The energy density of MHDE is
\begin{eqnarray}\nonumber
\rho_{\Lambda}=3c^{2}\pi^{2}L^{2}=\frac{3c^{2}\pi^{2}}{H^{2}}
\end{eqnarray}
which leads to
\begin{equation}\nonumber
\Omega_{\Lambda}=\frac{c^{2}\pi^{2}}{2H^{4}},
\end{equation}
and hence
\begin{equation*}
\dot{\Omega}_{\Lambda}=-2c^{2}\pi^{2}\frac{\dot{H}}{H^{5}}.
\end{equation*}
Replacing $\frac{\dot{H}}{H^5}$ and simplifying, it follows that
\begin{equation}\label{15a}
\dot{\Omega}_{\Lambda}=2c^{2}\pi^{2}(\frac{\dot{a}^{3}}{a^{3}}
-\frac{a^{4}\ddot{a}}{\dot{a}^5}).
\end{equation}
Now
\begin{equation}\label{17}
\frac{d\Omega_{\Lambda}}{da}=\frac{1}{Ha}\frac{d\Omega_{\Lambda}}{dt}
=2c^{2}\pi^{2}[\frac{\dot{a}^{2}}{a^{3}}-\frac{\ddot{a}a^{4}}{\dot{a}^{6}}].
\end{equation}
This is the equation of evolution of the modified holographic dark energy.

\section{Generalized Second Law of Thermodynamics}

The discovery of black hole thermodynamics has led to the
thermodynamics of cosmological models. Bekenstein \cite{47} proved
that there is a relation between an event horizon and thermodynamics
of black hole. The event horizon of black hole is in fact the
measure of entropy which is generalized to the cosmological models
so that each horizon corresponds to an entropy. The generalized
second law of thermodynamics is generalized in such a way that the
sum of time derivative of each entropy must be increasing. Since
horizon is a function of time, so the change in horizon causes
change in volume with respect to time. Consequently, energy and
entropy also change and hence both states have the common source
$T_{\mu\nu}$. We assume that temperature and pressure (which are the
fundamental ingredients for the discussion of GSLT) remain the same.

Here, we explore the validity of the generalized second law of
thermodynamics in the KK universe in which MHDE is interacting with
DM. The first law of thermodynamics is
\begin{equation}\nonumber
dS=\frac{PdV+dE}{T},
\end{equation}
where $T,~S,~E$ and $P$ are temperature, entropy, internal energy
and pressure of the system respectively. The corresponding entropies
for MHDE and DM will become
\begin{equation}\label{19}
dS_{\Lambda}=\frac{P_{\Lambda}dV +dE_{\Lambda}}{T},\quad
dS_{m}=\frac{P_{m}dV +dE_{m}}{T}.
\end{equation}
We discuss the GSLT when the the system is in equilibrium. In this
case, the temperature of the fluid and horizon are the same. We
avoid the concepts like chemical potential. The temperature and
entropy of horizon are defined as \cite{37c}
\begin{equation*}
T=\frac{1}{2\pi r_{a}},\quad
S_{h}=\frac{A}{4G}.
\end{equation*}
In this FRW type KK cosmology, we have
\begin{equation}\label{20}
T=\frac{1}{2\pi r_{A}}=\frac{1}{2\pi L},
\end{equation}
which corresponds to temperature of de-Sitter horizon and apparent
horizon in flat FRW cosmology \cite{37c}. Thus, this definition of
temperature and entropy of horizon work well in the FRW type KK
universe. From Eq.(\ref{4a}), we have
\begin{equation}\label{20b}
L^{-2}= H^{2}=\frac{1}{6}(\rho_{\Lambda}+\rho_{m}).
\end{equation}
The volume of the system is
\begin{equation}\label{21}
V=\frac{\pi^{2} L^{4}}{2}.
\end{equation}

Thermodynamical quantities are related to the cosmological
quantities by the following relations
\begin{eqnarray}
P_{\Lambda}=\omega^{eff}_{\Lambda}\rho_{\Lambda},\quad\label{22}
P_{m}=\omega^{eff}_{m}\rho_{m},\quad
E_{\Lambda}=\frac{\pi^{2}
L^{4}\rho_{\Lambda}}{2},\quad\label{23} E_{m}=\frac{\pi^{2}
L^{4}\rho_{m}}{2}.
\end{eqnarray}
In four dimension, $S_{h}=\frac{2\pi^{2}L^{3}}{4G}=4\pi^{3}L^{3}$
as $8\pi G=1$ leads to
\begin{equation}\label{24}
\dot{S_{h}}=12\pi^{3}L^{2}\dot{L}.
\end{equation}
Also, the time derivative of Eq.(\ref{19}) yields
\begin{equation}\label{25}
\dot{S_{\Lambda}}=\frac{P_{\Lambda}\dot{V}+\dot{E_{\Lambda}}}{T},\quad
\dot{S_{m}}=\frac{P_{m}\dot{V}+\dot{E_{m}}}{T}.
\end{equation}
From Eqs.(\ref{20}), (\ref{21})-(\ref{25}), we get
\begin{eqnarray}\label{26}
\dot{S}_{total}=
4\pi^{3}L^{4}[(1+\omega^{eff}_{\Lambda})\rho_{\Lambda}
+(1+\omega^{eff}_{m})\rho_{m}](\dot{L}-LH)+12\pi^{3}L^{2}\dot{L}.
\end{eqnarray}
where $S_{total}$ is the sum of three entropies. It follows from
Eqs.(\ref{11a})-(\ref{11b}) that
\begin{equation}\label{27}
(1+\omega^{eff}_{\Lambda})\rho_{\Lambda}+(1+\omega^{eff}_{m})\rho_{m}
=(1+\omega_{\Lambda})\rho_{\Lambda}+(1+\omega_{m})\rho_{m}.
\end{equation}
Using Eqs.(\ref{6a}), (\ref{7a}) and (\ref{20b}), it turns out that
\begin{equation}\nonumber
\dot{L}=\frac{L^3H}{3}{[(1+\omega^{eff}_{\Lambda})\rho_{\Lambda}+(1+\omega^{eff}_{m})\rho_{m}]}.
\end{equation}
Substituting this value of $\dot{L}$ in Eq.(\ref{26}), we have
\begin{equation}\label{28}
\dot{S}_{total}=\frac{4\pi^{3}L^{6}}{3}[{(1+\omega_{\Lambda})\rho_{\Lambda}+(1+\omega_{m})\rho_{m}}]^{2}.
\end{equation}
This implies that the time derivative of normal entropy plus horizon
entropy is an increasing function. Hence the second law of
thermodynamics holds for all time.

\section{Discussion and Concluding Remarks}

In this paper, we have taken compact FRW type KK universe along
with MHDE interacting with DM. We have developed equation of state
parameter as well as equation of evolution for this scenario. We
conclude from the equation of evolution that its evolution depends
on the scale factor and its derivatives. The first two terms in Eq.(\ref{14}), i.e.,
$-\frac{1}{2}+\frac{\ddot{a}a}{2\dot{a}^{2}}$,  can be written
in terms of deceleration parameter $q$ as $\frac{-(1+q)}{2}$. This indicates that the equation of
state parameter of MHDE is an increasing function.
Since the universe has an accelerating expansion with $q\leq-1$, this
means that $q+1\leq0$ implying $\frac{-(1+q)}{2}\geq0$. Similarly,
consider the third term in Eq.(\ref{14}) involving the energy
densities. The coincidence problem says that matter density is
nearly equal to the DE density in the present era. It follows that
$(\Omega_{\Lambda}-\Omega_{m})\sim0$ which implies that
$\frac{3b(\Omega_{\Lambda}-\Omega_{m})}{4\Omega_\Lambda}\sim0$.
This means that equation of state parameter depends majorally on the first
two terms, hence it is an increasing.

Further, we have investigated that GSLT holds for all time and its
validity is independent of the interacting form, equations of
state parameters, fifth dimension and the geometry of the black
hole. This validity of GSLT is proved for all times with the same
results as in FRW universe \cite{48}. Finally, we would like to
mention here that GSLT also holds for non-compact KK type universe
in which the variation along fifth dimension is very small.

\end{document}